\def\rn{\noindent\parshape 2 0truecm 8.8truecm 0.3truecm 8.5truecm}
\def\nn#1 #2{#1, #2.}				
\def\nnn#1 #2 #3{#1, #2. #3.}			
\def\nnnn#1 #2 #3 #4{#1, #2. #3. #4.}		
\def\dualand{, \&\hbox{ }}				
\def\multiand{, \&\hbox{ }}				
\def\rfprep#1;#2;#3 {{\par\rn#1 #2, preprint #3\par}}

\def\rg#1;#2;#3;#4;#5;#6 {\par\rn#1 #2, {\it #3}, {\bf #4}, #5 (``#6'') \par}
\def\rf#1;#2;#3;#4;#5 {\par\rn#1 #2, {\it #3}, {\bf #4}, #5\par}
\def\rfbook#1;#2;#3;#4;#5 {{\frenchspacing\par\rn#1 #2, {\it #3} (#4: #5)\par}}
\def\rfproc#1;#2;#3;#4;#5;#6 {{\frenchspacing\par\rn#1 #2, in {\it #3}, ed. #4 (#5: #6)\par}}

\def\expec#1{\langle#1\rangle}

\def\etal{{\frenchspacing\it et al.}}
\def\ie{{\frenchspacing\it i.e.}}
\def\eg{{\frenchspacing\it e.g.}}
\def\etc{{\frenchspacing\it etc.}}

\def\rms{{\frenchspacing r.m.s.}}

\def\caption#1{{\footnotesize #1}\smallskip\smallskip}

\def\beq#1{\begin{equation}\label{#1}}
\def\eeq{\end{equation}}
\def\beqa#1{\begin{eqnarray}\label{#1}}
\def\eeqa{\end{eqnarray}}
\def\eq#1{equation~(\ref{#1})}
\def\eqn#1{~(\ref{#1})}

\def\spose#1{\hbox to 0pt{#1\hss}}
\def\simlt{\mathrel{\spose{\lower 3pt\hbox{$\mathchar"218$}}
     \raise 2.0pt\hbox{$\mathchar"13C$}}}
\def\simgt{\mathrel{\spose{\lower 3pt\hbox{$\mathchar"218$}}
     \raise 2.0pt\hbox{$\mathchar"13E$}}}
\def\simpropto{\mathrel{\spose{\lower 3pt\hbox{$\mathchar"218$}}
     \raise 2.0pt\hbox{$\propto$}}}

\def\K{{\rm K}}

\def\GHz{{\rm GHz}}

\def\C{{\bf C}}
\def\E{{\bf E}}
\def\F{{\bf F}}
\def\I{{\bf I}}
\def\L{{\bf\Lambda}}
\def\N{{\bf N}}
\def\NN{{\bf\Sigma}}
\def\Nfg{{\bf N}_{fg}}
\def\R{{\bf R}}
\def\SS{{\bf S}}
\def\W{{\bf W}}
\def\e{{\bf e}}
\def\w{{\bf w}}
\def\x{{\bf x}}
\def\y{{\bf y}}
\def\ii{^{(i)}}
\def\xt{\tilde x}
\def\vxt{\tilde\x}
\def\yi{\y\ii}
\def\ith{i^{th}}
\def\jth{j^{th}}
\def\sigf{\sigma_{fg}}
\def\sign{\sigma_{n}}
\def\coherence{\xi}
\def\pmean{{\bar\phi}}
\def\psdev{{\Delta\phi}}
\def\amean{{\bar\alpha}}
\def\asdev{{\Delta\alpha}}

\def\l{\ell}

\def\ed{\end{document}}

\documentstyle[emulateapj,epsf,rotate]{article}
\begin{document}

\submitted{Submitted to ApJ November 30, 1997, revised January 31}

\title{REMOVING REAL-WORLD FOREGROUNDS FROM CMB MAPS$^*$}

\author{
Max Tegmark\footnote{Institute for Advanced Study, Princeton, 
NJ 08540; max@ias.edu}$^,$\footnote{Hubble Fellow} 
}

\begin{abstract}
Most work on foreground removal has treated the case where the
frequency dependence of all components is perfectly known 
and independent of position.
In contrast, real-world foregrounds are generally
not perfectly correlated between frequencies, with the
spectral index varying slightly with position and 
(in the case of some radio sources) with time.
A method incorporating this complication in presented,
and illustrated with an application to the upcoming
satellite missions MAP and Planck. 
We find that even spectral index variations as small
as $\Delta\alpha\sim 0.1$ can have a substantial impact on
how channels should be combined and on attainable accuracy.
\end{abstract}

\section{INTRODUCTION}
 
\def\thefootnote{\fnsymbol{footnote}}
\footnotetext[1]{
{\it Available in color from}
{\bf h t t p://www.sns.ias.edu/$\tilde{~}$max/foregrounds.html}
}
\def\thefootnote{\arabic{footnote}}

Future Cosmic Microwave Background (CMB) experiments 
can measure many key cosmological parameters to
great precision (Jungman {\etal} 1996; 
Bond {\etal} 1997; Zaldarriaga {\etal} 1997) --- in principle.
To achieve this in practice, foreground contamination
must be removed with comparable accuracy.
Tegmark \& Efstathiou (1996, hereafter ``TE96''),
derived the foreground subtraction method that minimized the residual
variance from foregrounds and noise under the assumption that
the frequency dependence of all components was perfectly known 
and independent of position.
This method, which was independently derived by
Bouchet \& Gispert (unpublished), has now been extensively tested
with simulations (see {\eg} Bouchet {\etal} 1995; 
Bersanelli {\etal} 1996) where each frequency channel
was the appropriate linear combination of a simulated CMB map,
foreground templates such as the Haslam, DIRBE and IRAS maps,
radio sources, and random noise. The inversion was found 
to accurately recover the input maps even though the foreground 
templates exhibited strong non-Gaussianity.

To further improve such modeling, one must incorporate the
complication that real-world foregrounds are generally
not perfectly correlated between frequencies, with the
spectral index varying slightly with position 
and (in the case of some radio sources) with time.
Such spatial variations of the spectral index have been observed
for both dust (\eg, Reach {\etal} 1996; Schlegel {\etal} 1997) 
and synchrotron radiation (Banday \& Wolfendale 1991; Platania {\etal} 1997),
and are of course even more pronounced for point sources
(\eg, Francheschini {\etal} 1989, 1991; Toffolatti {\etal} 1997).
As we will see, neglect of this complication can cause
severe underestimates of the residual foreground level in the cleaned
CMB map. It can also produce foreground residuals substantially
higher than can be obtained with the method we derive below.


\section{METHOD}


As in TE96, we assume that we have sky maps at $m$ 
frequencies $\nu_1,...,\nu_m$
(these maps may be internal channels of a CMB experiment, but can
include external templates such as the DIRBE maps as well), and
that these maps receive contributions from $n$ different physical
components (CMB, dust, \etc).

\subsection{Pixel by pixel or wavelet by wavelet?}

The general problem treated in this paper is 
how to take linear combinations of these $m$ maps 
to produce accurate maps of individual components.
The traditional approach (\eg, Brandt {\etal} 1994) has been 
to perform this multifrequency subtraction
separately for each pixel (direction in the sky).
However, this does not take advantage of the differences in 
smoothness between CMB and the various foregrounds.
The correlation between neighboring pixels is typically
stronger for diffuse galactic foregrounds (dust, synchrotron 
and free-free emission) than for CMB, where it is in turn stronger 
than for point sources.
One can therefore do better by performing the
linear combinations mode by mode rather than pixel by pixel,
using some variant of a Fourier expansion of the maps
(TE96). The optimal weights for the linear combination then differ
on large angular scales (where diffuse foregrounds are important) and 
small angular scales (dominated by point sources), as illustrated
in Figure 1. 
In addition to this scale dependence, there is also a direction
dependence,  since some pixels typically
have higher average levels of noise (from receiving less
observing time) or foreground contamination 
(from being closer to the galactic plane, say), 
than others. The suggestion of this author is therefore that
foreground subtraction be performed using modes that are
fairly localized both in real space and in Fourier space, for instance
some form of wavelets.

\subsection{Notation}
 
   All the methods discussed below can be applied 
regardless of which of the above-mentioned approaches is taken.
Let $y_j$ denote the is temperature $\delta T$ measured at 
the $\jth$ frequency in a given direction 
(or in a given mode --- in that case, $y_j$ is simply the corresponding 
multipole, Fourier, or wavelet coefficient).
Let $y\ii_j$ denote the contribution to $y_j$
from the $\ith$ physical component.
Grouping the measurements $y_j$ into an $m$-dimensional vector $\y$, 
we can thus write
\beq{yDefEq}
\y = \sum_{i=0}^n \yi.
\eeq
As in TE96, we assume that the different components have zero
mean ($\expec{\yi}=0$)\footnote{Another advantage of working 
with modes rather than pixels is that it can eliminate the nuisance of 
a non-zero mean (most foregrounds are strictly non-negative).
In a Fourier or multipole expansion, all modes but the (irrelevant)
monopole will have a vanishing average.
} 
and are uncorrelated, which means that
the data covariance matrix is simply given by
\beq{CdefEq}
\C\equiv\expec{\y\y^t} = \sum_{i=0}^n \C\ii,
\eeq
where $\C\ii\equiv\expec{\y\ii{\y\ii}^t}$ is the covariance matrix
of the $\ith$ component. It is convenient to factor this as
$\C\ii_{jk}=\R\ii_{jk}\sigma\ii_j\sigma\ii_k$, where the 
standard deviation and correlation is defined by 
$\sigma\ii_j\equiv [\C\ii_{jj}]^{1/2}$ and
$\R\ii_{jk}=\C\ii_{jk}/\sigma\ii_j\sigma\ii_k$, respectively.
For definiteness, let us take component $0$ to be the CMB.
Since the CMB temperature is the same in all channels,
equal to $x$, say, we thus have
$\y^{(0)}=\e x$, where the constant vector $\e$ is 
defined by $e_j=1$. We therefore have 
$\sigma^{(0)}_j=\expec{x^2}^{1/2}$, independent of $j$,
and the correlation matrix $\R^{(0)}=\E$, where $\E\equiv\e\e^t$,
a matrix consisting entirely of ones.
Let us take component $1$ to be the instrumental noise.
Then $\sigma^{(1)}_j$ is simply the {\rms} noise level in the
$\jth$ channel, and if the noise is uncorrelated between channels, 
we have $\R^{(1)}=\I$, the identity matrix. The remaining 
components (the various foregrounds) will typically have
correlation matrices $\R\ii$ that are intermediate between 
these two extreme cases of perfect correlation ($\R=\E$)
and no correlation ($\R=\I$).

Tegmark (1997, hereafter ``T97'') compared ten 
different methods for making CMB maps from time-ordered data.
The CMB foreground removal problem is quite analogous to the
mapmaking problem in that one seeks a linear 
inversion given certain assumptions about the ``noise''.
Indeed, all of the inversion methods described in T97 
can be used for foreground removal as well, and we will repeatedly return
to these connections below.

\subsection{A signal-to-noise eigenvalue problem}

Let us consider an arbitrary linear combination of the channels,
\beq{xtDefEq}
\xt\equiv\w\cdot\y,
\eeq
specified by some $m$-dimensonal weight vector $\w$.
If we want $\xt$ to estimate the $\ith$ component,
then $\y\ii$ is our signal and all the other components
act as noise. Let $\N$ denote the covariance matrix of this
generalized ``noise'', \ie, 
\beq{NdefEq}
\N\equiv\sum_{i\ne j}^n \C\ii.
\eeq
The contribution to the variance $\expec{\xt^2}$ of our estimator 
$\xt$ from
signal and noise is $\w^t\C\ii\w$ and $\w^t\N\w$, respectively.
Maximing the signal-to-noise ratio 
(maximing $\w^t\C\ii\w$ with $\w^t\N\w$ held fixed),
we find that $\w$ is a solution to the
generalized eigenvalue problem
\beq{EigenEq}
\C\ii\w=\lambda\N\w.
\eeq
This is analogous to the signal-to-noise eigenmode 
method (Bond 1995; Bunn \& Sugiyama 1995; Tegmark {\etal} 1997)
used in CMB power spectrum analysis, except that the data set $\y$
is now the measurement at different frequencies rather than 
at different positions in the sky. The $m$ different 
eigenvectors $\w$
give $m$ uncorrelated estimators $\xt$, the least noisy one
being that corresponding to the largest eigenvalue $\lambda$. 

\subsection{TE96 as a special case}

Throughout the rest of this paper, we limit our attention to 
estimating component $0$, the CMB.
Since $\C^{(0)}\propto\E$, a matrix of rank 1 (with only one
non-zero eigenvalue, which corresponds to the eigenvector $\e$),
the eigenvalue problem reduces to a simple matrix inversion for this
case:
\eq{EigenEq} gives $\N\w\propto\E\w=\e(\e^t\w)\propto\e$,
so $\w\propto\N^{-1}\e$.
Normalizing $\w$ so that $\w^t\E\w=1$, we obtain
\beq{wEq}
\w = {\N^{-1}\e\over\e^t\N^{-1}\e}.
\eeq
This normalization corresponds to
$\sum w_i=\e\cdot\w=1$, so we can interpret $\xt$ as simply a weighted
average of the $m$ channels, with $w$ giving the weights (some weights
may be negative).

In TE96, we assumed that the frequency dependence 
of each component was independent of position and time. 
Since this means that the map of a component looks the same
at all frequencies, apart from an overall frequency-dependent 
scale factor, this assumption is equivalent to saying that 
each component is perfectly correlated between frequencies, 
\ie, that $\R\ii=\E$ except for $i=1$, the instrumental noise
component. 
Following the notation of TE96, let $\SS$ denote the diagonal
covariance 
matrix of the different components at some 
fiducial frequency $\nu_*$ (say 100 \GHz), and let
$F_{ji}$ specify the {\rms} of the $\ith$ component
at the $\jth$ frequency relative to the value at $\nu_*$.
The correspondence between TE96 and our equations is then given 
by $\sigma\ii_j = F_{ji} S_{ii}^{1/2}$.
Defining $\NN\equiv\C^{(1)}$ and $\Nfg\equiv\sum_{i=2}^n\C\ii$
as the contributions to the covariance matrix 
$\N$ from receiver noise and foregrounds, respectively, 
we can thus write
\beq{TE96Neq}
\N = \Nfg + \NN = \F\SS\F^t + \NN,
\eeq
TE96 reconstructed all components, not merely the CMB,
with an estimator of the form $\vxt=\W\y$.
Since $\R\ii=\E$ for all the foregrounds, \eq{EigenEq} will give a 
single eigenvector with $\lambda>0$ for estimating each one, 
just as for the CMB component.
Arranging these vectors $\w$ as the rows of the matrix $\W$
and performing the relevant algebra, we obtain
\beq{TE96eq}
\W = \L\F^t[\F\SS\F^t+\NN]^{-1},
\eeq
with $\Lambda_{jk}\equiv\delta_{jk}/(\W\F)_{jj}$,
\ie, equation (36) of TE96. 
We have thus generalized the TE96 result, and found
that it corresponds to the special case of 
perfect foreground correlations, $\R\ii=\E$.
Conversely, it is easy to show that our method\eqn{wEq}
can be derived  from the TE96 method by replacing each foreground
component with many sub-components, each with a slightly
different frequency dependence, as described in \S 5.4 of
TE96.

\section{THE TRADEOFF BETWEEN FOREGROUNDS AND NOISE}

Our discussion above has illustrated that foregrounds
are very much like detector noise --- they are simply 
more correlated between channels. 
When chosing $\w$ to make
a CMB map, there is generally a tradeoff between 
the amount of residual noise 
$\sign^2\equiv[\w^t\NN\w]^{1/2}$ and residual 
foreground contamination 
$\sigf^2\equiv[\w^t\Nfg\w]^{1/2}$.
This is clearly seen if we minimize 
$\sigf^2$ for some fixed level of noise $\sigf^2$,
maintaining our normalization constraint $\e\cdot\w=1$.
Solving this constrained minimization problem by introducing
Lagrange multipliers $\gamma$ and $\lambda$, 
this corresponds to minimizing the expression
$\w^t[\Nfg+\gamma\NN]\w - \lambda\e\cdot\w$,
which gives $\w\propto[\Nfg+\gamma\NN]^{-1}\e$.
We recognize this as our solution in \eq{wEq}, but with
$\N=\Nfg+\NN$ replaced by $\Nfg+\gamma\NN$, {\ie}, with the noise level
rescaled relative to its true value by a factor $\gamma$.
For the TE96 case, this gives 
\beq{TE97eq}
\W = \L\F^t[\F\SS\F^t+\gamma\NN]^{-1},
\eeq
which corresponds to ``Method 8''
in the method table of T97.
Below we will see that this free parameter $\gamma$ can be chosen to
indicate how concerned we are about $\sign$ relative
to $\sigf$.  

\begin{center}
{\footnotesize
\begin{tabular}{||l||c|c|c|c||}
\hline
Channel spec.           &COBE	&MAP    &LFI	&HFI\\
\hline
$\nu$ [GHz]             &	&22     &       &100    \\
FWHM [arcmin]           &	&57     &       &10.6   \\
$10^6 \Delta T/T$       &	&10     &       &1.81   \\
\hline
$\nu$ [GHz]             &31.5	&30     &30  	&143    \\
FWHM [arcmin]           &425	&41     &33	&7.4    \\
$10^6 \Delta T/T$       &77	&12	&1.6    &2.1    \\
\hline
$\nu$ [GHz]             &51	&40     &44	&217    \\
FWHM [arcmin]           &425	&28     &23  	&4.9    \\
$10^6 \Delta T/T$       &31	&10     &2.4 	&4.6    \\
\hline
$\nu$ [GHz]             &	&60     &70  	&353    \\
FWHM [arcmin]           &	&21     &14  	&4.5    \\
$10^6 \Delta T/T$       &	&13     &3.6 	&15.0   \\
\hline
$\nu$ [GHz]             &90     &90     &100 	&545    \\
FWHM [arcmin]           &425	&13     &10  	&4.5    \\
$10^6 \Delta T/T$       &34	&13     &4.3 	&144.0  \\
\hline
$\nu$ [GHz]             &       &       &       &857    \\
FWHM [arcmin]           &       &       &       &4.5    \\
$10^6 \Delta T/T$       &       &       &       &4630   \\
\hline
\end{tabular}
}
\end{center}
{\footnotesize
{\bf Table  1.} ---  Specifications used for COBE, MAP and Planck.
}
\bigskip

Figure 2 shows the result of applying \eq{TE97eq}
to the three satellite experiments COBE, MAP and Planck, with
the lines corresponding to $\gamma$ ranging from 
$0$ to $\infty$. 
Here we have used four foreground components ($n=5$):
dust, free-free emission, synchrotron radiation and point sources.
These are modeled as in TE96, with some minor updates as shown 
on Figure 1 to reflect recent foreground measurements
(Bersanelli {\etal} 1996; Kogut {\etal} 1996ab; de Oliveira-Costa {\etal} 1997; 
Toffolatti {\etal} 1997).
We have used the experimental specifications shown in Table 1,
taken from Bennett {\etal} (1996) and the MAP and Planck
web sites ({\it http://map.gsfc.nasa.gov}
and 
{\it http://tonno.tesre.bo.cnr.it/research/planck/tab\_sens.htm}).
LFI and HFI refers to the low and high frequency instruments on board
Planck.

Our derivation above showed that no method can
give a point $(\sign,\sigf)$ below or to the left of this line, 
{\ie}, that \eq{TE97eq} minimizes the foreground residual for
any given noise level $\sigf$.
The original TE96 method ($\gamma=1$, indicated by
a solid square) corresponds to minimizing 
the {\it total} residual variance, {\ie}, $\sign^2+\sigf^2$,
so in a linear-linear plot, the TE96 point lies 
where the line is closest to the origin.
As we increase $\gamma$, the algorithm cares more about reducing 
noise and less about foregrounds. The upper endpoint of the line, with
$\gamma=\infty$, corresponds to $\w\propto\NN^{-1}\e$,
which if the detector noise is uncorrelated ($\R^{(1)}=\I$)
is a simple minimum-variance weighting, ignoring the foregrounds.
For COBE, this extreme case is seen to minimize the total 
residuals as well --- indeed, most published analyses of the COBE data
made this choice, abstaining from foreground subtraction.

\centerline{{\vbox{\epsfxsize=8.5cm\epsfbox{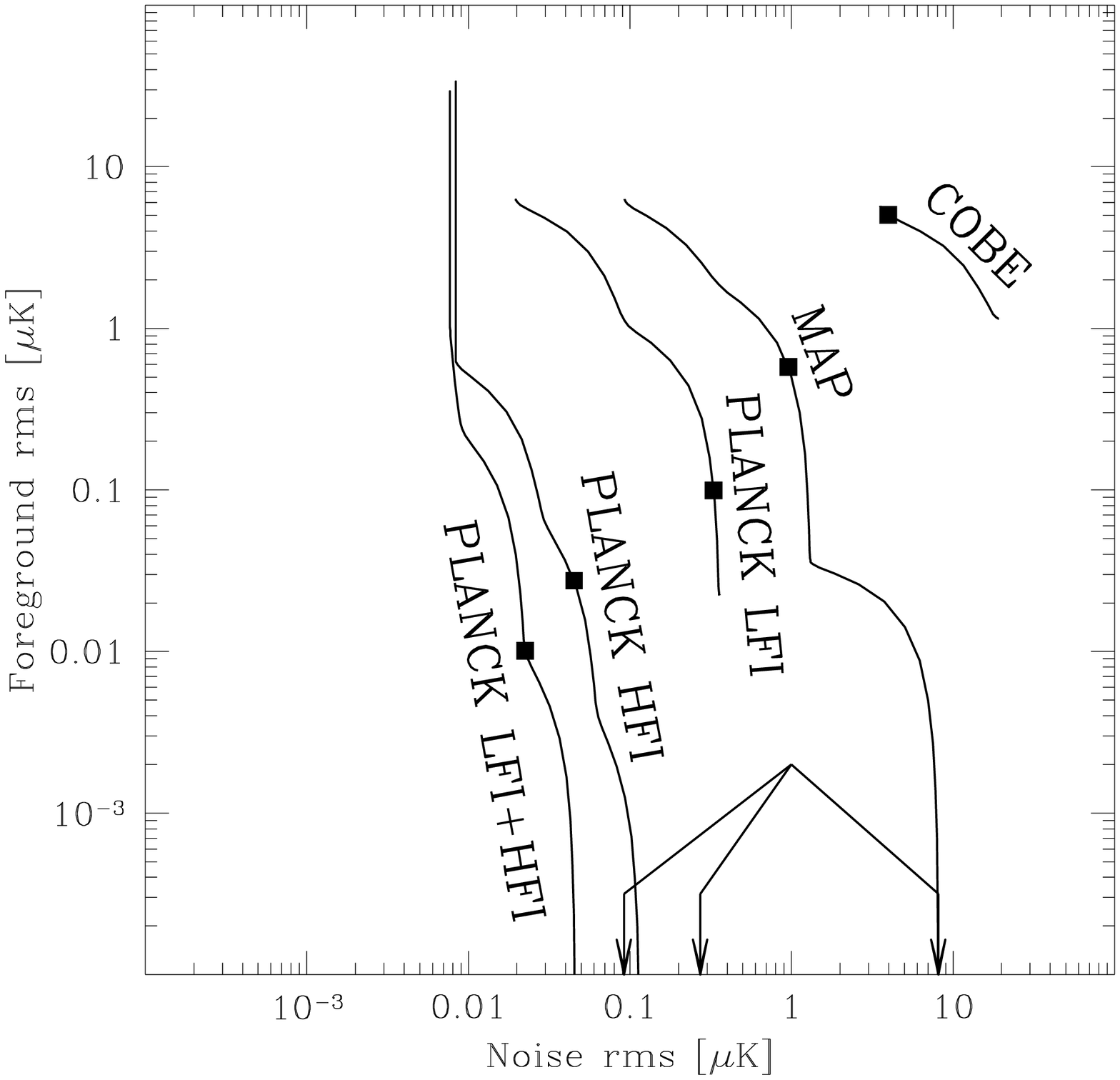}}}}
\vskip-0.1cm
\caption{
{\bf Figure 2.} The curves show the smallest residual foreground level 
attainable for a given noise level, assuming that the
frequency dependence of the foregrounds is perfectly known.
The total {\rms} residual $(\sign^2+\sigf^2)^{1/2}$ is minimized
at the solid squares (the TE96 method). Arrows point to $(\sign,\sigf)$
for the marginalization method, which is seen to give 
$\sigf=0$ (off the scale), but occasionally at a higher noise cost 
than necessary.
These curves are for a mode
$a_{\ell m}$ with $\l=10$ --- due to their differences in 
angular resolution, the experiments differ more 
dramatically for larger $\l$ as well as on a pixel-by-pixel basis.
}

There may be good reasons to be more concerned about foregrounds 
than detector noise. For instance, they tend to be
non-Gaussian and we are usually unable to model
their frequency and scale dependence 
as accurately as for detector noise. If we in this vein
decrease $\gamma$, we move downward along the curve. 
For cases when the number of channels equals or exceeds
the number of components (as for HFI and MAP),
complete foreground removal is possible: $\sigf\to 0$
as $\gamma\to 0$, corresponding to a weighting where $\w$
is the first row of $[\F^t\NN^{-1}\F]^{-1}\F^t\NN^{-1}$
(``Method 3'' of T97). For these cases, the factor by which
$\sign$ increases as we go from no foreground removal 
(upper endpoint) to complete foreground removal
(lower endpoint) is the {\it Foreground Degradation Faction} (FDF)
introduced by Dodelson (1996).  
The residual noise is also shown (by arrows at bottom)
for the marginalization method derived by Dodelson (1996), 
in which $\w$ is the first row of $[\F^t\F]^{-1}\F^t$
(``Method 2'' in T97). This is seen to give
an FDF that is about a factor of 2 larger for the HFI (m=6) and
HFI-LFI (m=10) cases, but identical to the TE96 method for MAP (m=5).
The reason is that when 
there are more channels than components
($m>n$), there are $m-n$ degrees of freedom left in $\w$ after we
have required that the foregrounds be eliminated and imposed the
normalization constraint. The TE96 formula uses these extra 
degrees of freedom to minimize $\sign$. 

Although the above-mentioned reasons for trying to 
reduce $\sigf$ below $\sign$ 
(which might require $\gamma <1$) may be valid, they are
not grounds for outright foreground paranoia.
Attempts to push $\sigf$ down say a factor of ten below 
$\sign$ are probably overkill and not worth the heavy cost in terms 
of increased noise. 
Most importantly, as we will see in the next section, such attempts 
are likely to be misleading, since 
even tiny departures from perfect correlations can
reintroduce non-negligible foreground residuals.
Although the $\gamma=0$ method has been advocated as
conservative (Dodelson \& Stebbins 1994; Dodelson 1996), 
since it requires no assumptions about the 
{\it amplitude} of the foreground fluctuations, we will see that
it is quite sensitive to assumptions about their frequency dependence.

\section{THE EFFECT OF FREQUENCY COHERENCE}

Figure 2 assumed perfect foreground correlations, $\R\ii=\E$ for $i>1$.
We will now relax this assumption. 

\subsection{A toy model}

To illustrate the qualitative changes
that occur, let us derive 
a simple toy model in which we can relate the correlation
matrices $\R\ii$ to more familiar quantities.
Given some foreground component $i$ and two frequencies
$\nu_j$ and $\nu_k$, we define 
$\phi_-\equiv y\ii_j$, $\phi_+\equiv y\ii_k$,
$\phi\equiv(\phi_-\phi_+)^{1/2}$, $\eta\equiv\nu_k/\nu_j$
and $\alpha\equiv\ln(\phi_+/\phi_-)/\ln\eta.$ 
Thus $\phi_-$ and $\phi_+$ denote the brightness of a
pixel at the two frequencies, $\phi$ is the (geometric)
mean brightness, and $\alpha$, the
``color'', is the spectral index for which a power law spectrum
$\phi(\nu)\propto\nu^\alpha$ would connect $\phi_-$ with $\phi_+$.
With this notation, we have
\beq{phiEq}
\phi_\pm=\phi\eta^{\pm\alpha/2}.
\eeq
Let us make the simplifying assumption that
the brightness $\phi$ and the color $\alpha$ are statistically
independent. Although probably not very accurate, this 
approximation is motivated by the fact that 
$\phi$ depends strongly on color-independent quantities
such as the distance (in the case of radio sources) and
on the amount of emitting material along the line of sight
(in the case of the diffuse foreground components).
Using this independence gives
\beqa{ExpecEq}
\expec{\phi_\pm}&=&\expec{\phi}\expec{\eta^{\pm\alpha/2}},\\
\expec{\phi_\pm^2}&=&\expec{\phi^2}\expec{\eta^{\pm\alpha}},\label{ExpecEq2}\\
\expec{\phi_-\phi_+}&=&\expec{\phi^2}\label{ExpecEq3}.
\eeqa
We define the means and standard deviations
$\amean\equiv\expec{\alpha}$,
$\pmean\equiv\expec{\phi}$,
$\asdev\equiv(\expec{\alpha^2}-\amean^2)^{1/2}$, 
$\psdev\equiv(\expec{\phi^2}-\pmean^2)^{1/2}$.
Let us also assume that the quantity 
$\asdev\ln\eta\ll 1$, so that a fairly 
definite spectral index will be apparent in a scatter plot
of $\ln\phi_+$ against $\ln\phi_-$. 
Taylor expanding the exponential, 
this allows us to make the approximations
$\expec{\eta^{\pm\alpha}}
=\eta^{\pm\amean}\expec{e^{\pm(\alpha-\amean)\ln\eta}}
\approx\eta^{\pm\amean}e^{(\asdev\ln\eta)^2/2}$ 
and
$\expec{\eta^{\pm\alpha/2}}
\approx\eta^{\pm\amean/2}e^{(\asdev\ln\eta)^2/8}$.
Substituting this into 
equations\eqn{ExpecEq}$-$(\ref{ExpecEq3}), we can compute
the standard deviations 
$\Delta\phi_\pm\equiv(\expec{\phi_\pm^2}-\expec{\phi_\pm}^2)^{1/2}$
and the correlation.
We find that
$\phi_+/\phi_-\approx\Delta\phi_+/\Delta\phi_-\approx\eta^\amean$,
so the mean brightness
and the {\rms} fluctuations scale in the same way with frequency,
as expected.
The correlation coefficient is given by
\beq{RmodelEq}
\R\ii_{jk}\equiv 
{\expec{\phi_-\phi_+}-\expec{\phi_-}\expec{\phi_+}
\over\Delta\phi_+\Delta\phi_-}
\approx e^{-(\ln\nu_j-\ln\nu_k)^2/2\coherence^2},
\eeq
where 
\beq{CoherenceDefEq}
\coherence\equiv {1\over\asdev\left(1+\beta^2\right)^{1/2}}.
\eeq
and $\beta\equiv\pmean/\psdev$ is the ratio of the mean brightness
to the {\rms} fluctuations.
We will call the parameter $\coherence$ the
{\it frequency coherence}, since it determines how many powers of 
$e$ we can change the frequency by before the correlation starts
breaking down. The two limits $\coherence\to 0$ and 
$\coherence\to\infty$ correspond to the two extreme
cases $\R\ii=\I$ and $\R\ii=\E$ that we encountered above.
Since the temperature in a foreground map typically 
range from its maximum down to values near zero, with
the {\rms} fluctuations $\psdev$ being of the same order
of magnitude as the mean $\pmean$, $\beta$ is usually of
order unity and we arrive 
at the following useful rule of thumb: 
{\it The frequency coherence is of the order of the inverse
spectral index dispersion}.

\centerline{{\vbox{\epsfxsize=8.5cm\epsfbox{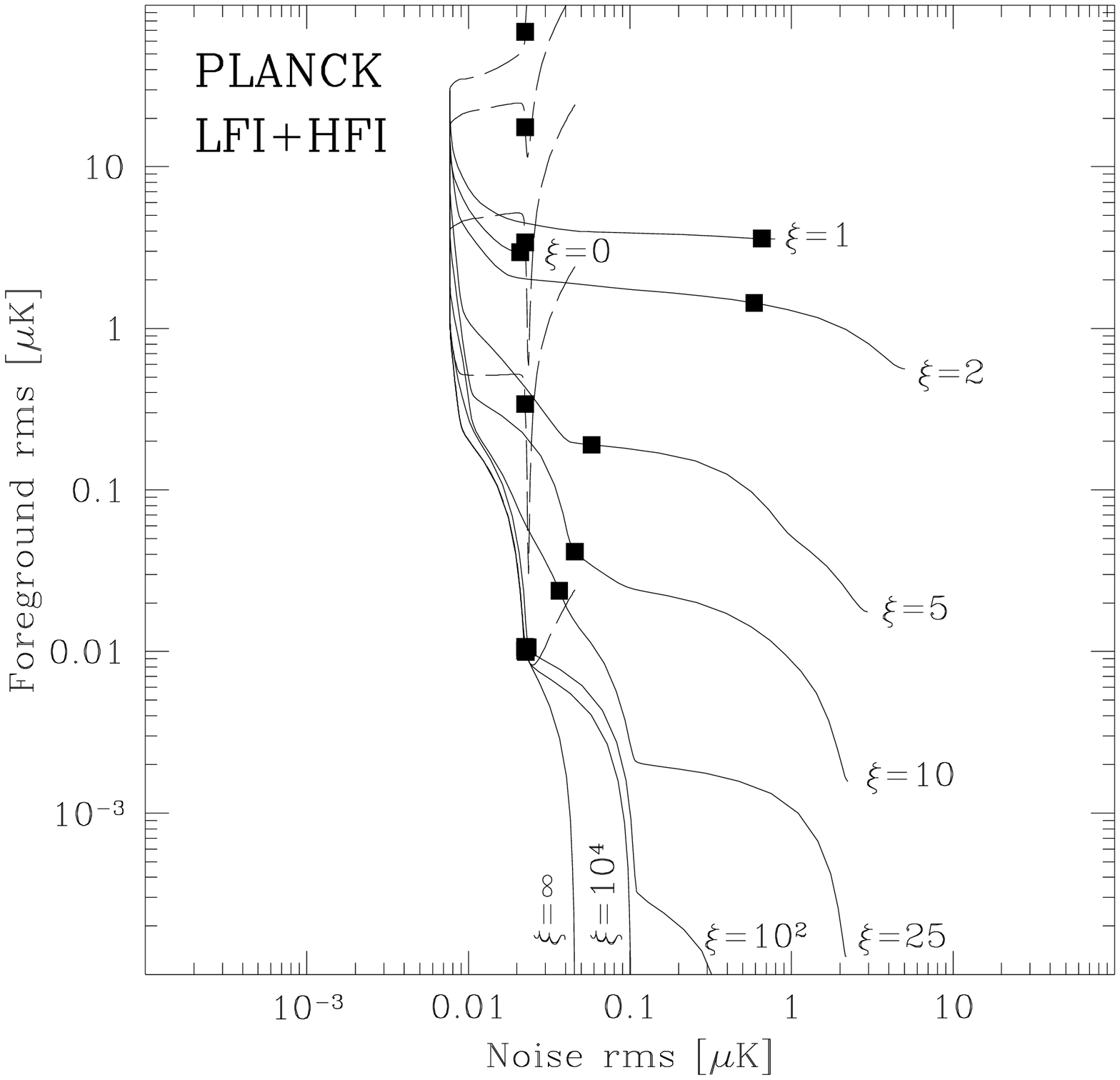}}}}
\vskip-0.1cm
\caption{
{\bf Figure 3.} Same as figure 2, but for Planck HFI+LFI only,
varying the frequency coherence $\coherence$.
The five dashed (unlabeled) 
curves show the result of assuming $\coherence=\infty$
when in fact $\coherence=$ 0, 2, 10, 100 and $10^4$
(from top to bottom).
}

Figure 3 shows how the Planck results from figure 2 
($\coherence=\infty$) change when $\coherence$ is reduced.
The solid squares correspond to \eq{wEq}, and the tradeoff
curves are generated by rescaling the receiver noise contribution
to $\N$ in \eq{wEq} by different constants.
We see that these curves follow 
the $\coherence=\infty$ curve down from the top, then branch off to 
the right at a foreground level that depends on $\coherence$.
To reduce the foreground residual below this level becomes 
extremely costly in terms of extra noise 
(giving a large FDF, in Dodelson's terminology). 
Complete foreground removal is of course
impossible when $\coherence<\infty$.
The dashed curves show that if the subtraction method assumes
ideal ($\coherence=\infty$) foregrounds, it is disastrous to 
be too greedy and try to push the $\sigf$ way below $\sign$, 
since this can actually make things worse!

Note that since we assumed that $\asdev\ln\eta\ll 1$, 
our derivation of \eq{RmodelEq} only applies to the 
first 3 terms in a Taylor expansion, showing that 
$R\ii_{jk}=f(\ln\eta/\coherence)$ where $f(x) = 1-x^2/2+...$.
We recomputed Figure 3 for a variety of such function of 
the form $f(x) = (1+x^2/2n)^{-n}$ ($n=\infty$ gives
the Gaussian of \eq{RmodelEq}, $n=1$ gives a Lorentzian,  {\etc}),
and found that the shape of the far wings of $f$ is only of 
secondary importance --- the main question is
how correlated neighboring channels are, 
which for $\coherence\gg 1$ depends mainly on
the curvature of $f$ near the origin. Narrower 
wings (larger $n$) can occasionally help slightly, 
just as $\coherence=0$ is better than $\coherence=1$ in 
Figure 3.

\section{CONCLUSIONS}

When removing CMB foregrounds, one can take advantage 
of all ways in which they differ from CMB fluctuations.
\begin{enumerate}
\item 
Non-Gaussian behavior can be exploited to throw out
severely contaminated regions (\eg, bright point sources,
the Galactic plane).
\item Their frequency dependence can be exploited  
to subtract them out as we have described above.
\item Knowledge of their power spectra can be used by
including residual foreground fluctuation amplitudes
as additional free parameters when fitting the measured
power spectrum to theoretical models.
\end{enumerate}
The TE96 subtraction method (for step 2) 
has been shown (T97) to be lossless
(retain all the cosmological information) if the foregrounds
are Gaussian with $\coherence=\infty$, 
and if the subtraction is
performed mode by mode (as suggested 
by TE96 and implemented by Bouchet {\etal} 1995)
rather than pixel by pixel --- the
latter destroys information by not taking advantage of
correlations between neighboring pixels. 
In this {\it Letter}, we have studied the more realistic
case $\coherence<\infty$, and found that even  
spectral index variations as small as $\asdev=0.1$ make a
substantial difference for the choice of method and for attainable 
results.
Complete foreground removal becomes impossible, and 
attempting this nonetheless by assuming $\coherence=\infty$
can even be worse than no foreground removal at all.
It is easy to show that the method of \eq{wEq} is lossless 
with the same assumptions, for any $\coherence$.
This means that one more property needs to be determined
for each foreground component, in addition to its 
dependence on frequency and scale: its frequency correlations
$\R\ii$. 
With a simple toy model, we illustrated that this is directly 
linked to the spectral index dispersion $\asdev$.
$\asdev$ could easily be as large as 
0.1 for synchrotron radiation, 0.3 for dust, 0.01 for free-free emission
and 0.5 for radio sources if we neglect sources of prior information
about $\alpha$.
It has been argued that the spectral index for dust
depends on galactic latitude (\eg, Reach {\etal} 1995), 
whereas that for synchrotron emission 
is correlated with both the spectral index that can be measured at lower
frequencies (Brandt {\etal} 1995) and 
with the degree of synchrotron 
polarization (Bernstein 1992). 
By improving our understanding and modeling
of how the foreground spectral indices vary 
with position, it may thus be possible to
reduce the effective $\asdev$, thereby 
improving our foreground removal and the accuracy
with which cosmological parameters can be measured with the CMB. 

\bigskip
The author wishes to thank Ang\'elica de Oliveira-Costa, Scott Dodelson,
George Efstathiou, 
Lyman Page and David Wilkinson for useful discussions and suggestions.
This work was supported by
NASA grant NAG5-6034 and Hubble Fellowship
{\#}HF-01084.01-96A, awarded by the Space Telescope Science
Institute, which is operated by AURA, Inc. under NASA
contract NAS5-26555.


\clearpage
\onecolumn
\centerline{\epsfxsize=17cm\epsfbox{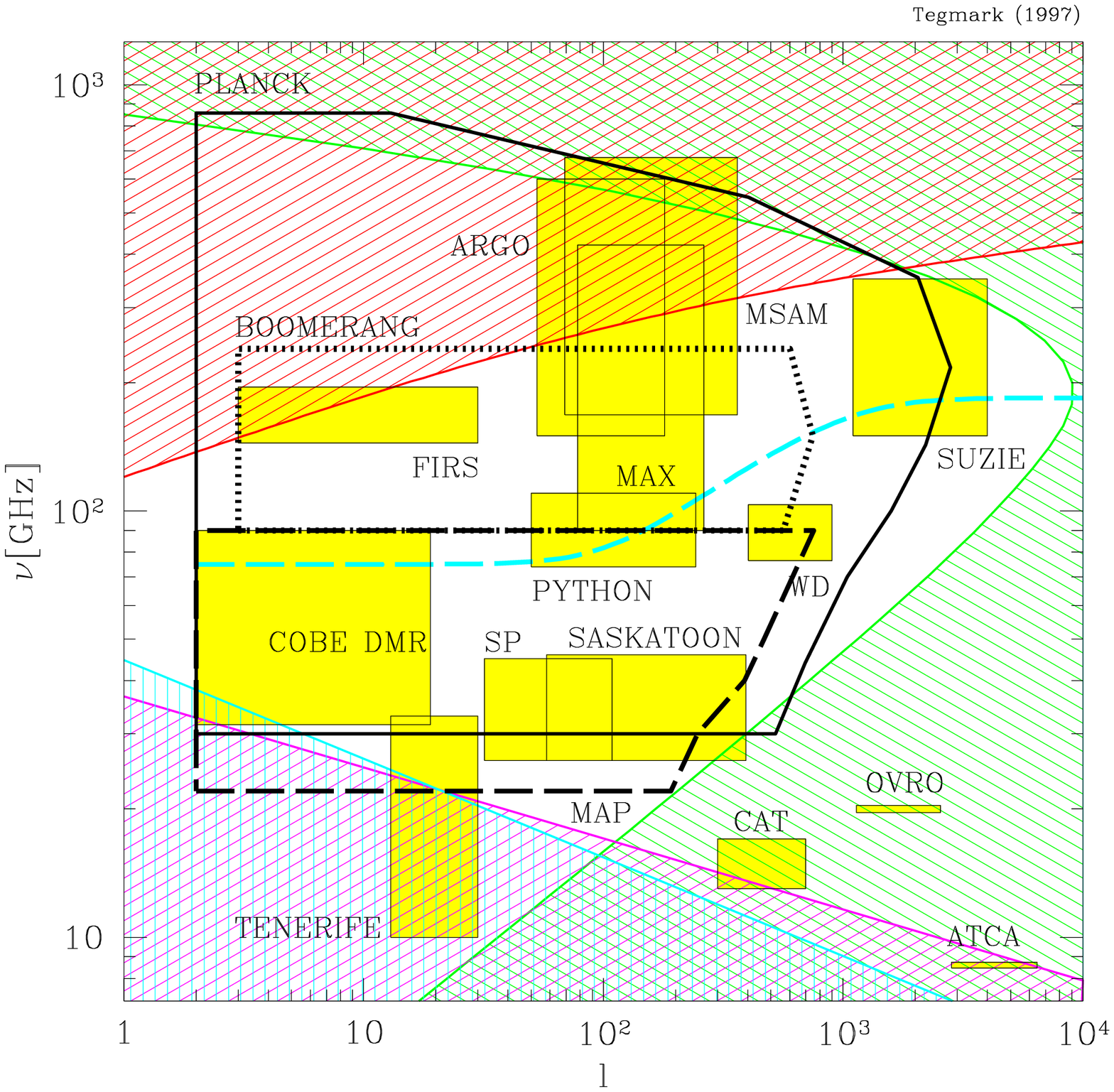}}
\caption{{\bf Figure 1.} Where various foregrounds dominate.
The shaded regions indicate where the various foregrounds cause
fluctuations exceeding those of COBE-normalized scale-invariant
fluctuations,
thus posing a substantial challenge
to estimation of genuine CMB fluctuations.
They correspond to dust (top), free-free emission (lower left),
synchrotron radiation (lower left, vertically shaded)
and point sources (lower and upper right).
The heavy dashed line shows the frequency where the total foreground
contribution to each multipole is minimal.
The boxes roughly indicate the range of multipoles $\l$ and frequencies
$\nu$ probed by various CMB experiments, as in TE96.
}

\end{document}